\documentstyle[preprint,prl,aps]{revtex}      
\begin{document}                             
\draft\tighten                               

\preprint{IUCF-UP-UWM-WMU}

\title{\Large\bf Dependence of $\rm \vec{p}\vec{p}\rightarrow pp\pi^{\circ}$ near Threshold\\ 
              on the Spin of the Colliding Nucleons}

\author {H.O. Meyer, J.T. Balewski, M. Dzemidzic, J. Doskow,
R.E. Pollock, B. v. Przewoski,\\ 
   T. Rinckel, F. Sperisen, P. Th\"orngren-Engblom, M. Wolanski}
\address{Dept. of Physics and Cyclotron Facility , Indiana University,
Bloomington, Indiana 47405}
\author{ W. Haeberli, 
B. Lorentz, F. Rathmann\footnote{present address:
   Forschungs Zentrum J\"ulich GmbH, 52425 J\"ulich, Germany.},
B. Schwartz, T. Wise}
\address{      University of Wisconsin-Madison, Madison, WI, 53706}
\author {W.W. Daehnick, R.W. Flammang,
Swapan K. Saha\footnote{permanent address:
   Bose Institute, Calcutta 700009, India}, D.J. Tedeschi\footnote{Physics Department, Univ. of
South Carolina, Columbia, SC 29208}}
\address{Dept. of Physics and Astronomy, University of Pittsburgh, Pittsburgh, PA 15260}
\author{P.V. Pancella}
\address       {Western Michigan University, Kalamazoo, MI, 49008}
                                
\date{\today}
\maketitle

\begin{abstract}                             
     A polarized internal atomic hydrogen target and a stored, polarized
beam are used to measure the spin-dependent total cross section
 $\rm \Delta\sigma_T/\sigma_{tot}$, as well as the polar integrals of the spin
correlation coefficient combination $\rm A_{xx}-A_{yy}$,
and the analyzing power $\rm A_y$ for $\rm \vec{p}\vec{p}\rightarrow pp\pi^{\circ}$ at four bombarding energies
between 325 and 400~MeV. This experiment 
is made possible by the use of a cooled beam in a storage ring.
 The polarization observables are used to study
the contribution from individual partial waves.
\end{abstract}



     Close to threshold, pion production in the nucleon-nucleon (NN)
system involves only a few partial waves, because the relative kinetic
energies between the particles in the final state are small. At the same time,
the transferred momentum is large, making this reaction sensitive to the
short-range NN interaction. This gives us the rare opportunity to study
the NN interaction in a regime where the nucleons start to overlap, while
being able to experimentally separate contributions from different angular
momentum states. For instance, in $\rm pp\rightarrow
pp\pi^{\circ}$, the transition from a $\rm ^3P_0$ initial
state to a $\rm ^1S_0$, $\rm l_{\pi}=0$ final state (or short, Ss) can be isolated by lowering the
bombarding energy until all higher partial waves have become
insignificant. This is the case for bombarding energies below about
300~MeV. 
A measurement of the total cross section, $\rm \sigma_{tot}$, in this 
regime\cite{R01,R02}
became possible with the advent of storage ring technology (for a review,
see\cite{R03}). At the time of that measurement, calculations using the ansatz of
Koltun\cite{R04} under-predicted the total cross section by about a factor of five.
Subsequently it was suggested that off-shell rescattering\cite{R05}, and an
enhancement of the axial charge from the exchange of heavy mesons 
($\rm \omega$, $\rm \sigma$)\cite{R06} are important processes which were previously omitted. First
attempts were also made to relate this process to fundamental symmetries
by use of chiral perturbation theory (Ref.\cite{R07}, and references therein).

    The present experiment  uses a polarized beam on a polarized target
to extend the study of the $\rm pp\rightarrow pp\pi^{\circ}$ reaction to the higher partial waves
with p-wave nucleons and s-wave pions (short, Ps), or p-wave nucleons
and pions (Pp) in the final channel (the combination Sp is forbidden). 
Bombarding energies were chosen to cover the range over which the higher
partial waves gradually become significant, and finally dominate the
reaction. Except for a few analyzing power data of low precision\cite{R08,R09}, this is
the first investigation of polarization observables in this reaction
below 
400~MeV.

     The experiment was carried out with the Indiana Cooler. Protons
from the cyclotron were stack-injected into the ring at 197~MeV, reaching
an orbiting current of several 100 $\rm \mu$A within a few minutes. The beam was
then accelerated to the energies listed in Table~1. After typically 10 minutes
of data taking, the remaining beam was discarded, and the cycle was
repeated. The beam polarization was vertical, either up or down; the 
direction was inverted after each cycle.

     Experience in the use of the polarized target and stored, polarized
beam was gained during a series of precise measurements of spin
correlation coefficients in pp elastic scattering\cite{R10}. The internal polarized
target is realized by injecting a beam of polarized atoms into an
open-ended 
storage cell (S in Fig.~1) which, for this experiment, consisted
of  
a 12~mm diameter aluminum tube of 25~$\rm \mu$m wall thickness. Such a cell
minimizes the amount of material close to the beam, and allows an
unobstructed view of the full azimuthal range. The cell is coated with
Teflon to avoid depolarization of atoms colliding with the wall. Its position
can be adjusted remotely to minimize beam halo intercepting the cell wall. 
During data taking, the target polarization is changed every 2~s pointing in
sequence, up, down, left, right. A target polarization change takes less than
100~ms and does not affect the magnitude of the polarization\cite{R10}.

     The detector arrangement is shown in Fig.~1. Reaction products from
the target region exit the vacuum chamber through a 0.18~mm thick
stainless steel foil (not shown). Two segmented scintillator arrays (E, K)
with a total thickness of 25~cm are designed to stop the protons from
$\rm pp\rightarrow pp\pi^{\circ}$ : if a particle reaches detector V, the event is vetoed. Outgoing
charged particles are identified by their energy and their time of flight
between the F and E scintillators. Their direction is measured by a set of
four wire planes (WC1, WC2). The detector setup is similar to the one used
for the $\rm pp\rightarrow pp\pi^{\circ}$ total cross section measurement\cite{R01}, with
some changes to cover the larger angular range at the higher energies of
this measurement. Typical distributions of the reconstructed mass,
$\rm m_x$, of the
undetected particle are shown in Fig.~2. The events of interest are those in
the $\rm \pi^{\circ}$ peak. The background under the peak (less than
10\% of the total yield) is estimated using a background shape which is
consistent with data obtained with only $\rm N_2$ gas in the target
cell.

     Proton-proton elastic scattering was used to monitor polarization and
luminosity, concurrently with the acquisition of $\rm \vec{p}\vec{p}\rightarrow pp\pi^{\circ}$ events. To this
end, coincidences between two protons exiting near  
$\rm \theta_{lab}=45^{\circ}$ are detected
by two pairs of scintillators placed at $\rm \Phi =\pm 45^{\circ}$ and
$\rm \Phi =\pm 135^{\circ}$ (X in Fig.~1).
For vertical beam and horizontal target polarization this arrangement
measures the spin correlation coefficient combination $\rm A_{xx}-A_{yy}$ which, for
pp scattering,  is large and known\cite{R11}. This yields an accurate
measurement of the product of beam and target polarization, 
$\rm P_{BT}\equiv P_B\cdot P_T$ (see
Table~1). In addition, the time-integrated luminosity,  $\rm \int$Ldt, is deduced from
the known pp elastic cross section (see Table~1). After correcting for
particles lost through the central hole  (discussed below), the
inefficiency of the wire chambers  ($\sim$7\%), and the loss of events due to reactions
in the scintillators (6-10\%), we verified that the observed number of
$\rm pp\pi^{\circ}$ 
events agrees  with the total cross section in the literature, ranging
from about 8~$\rm \mu$b at 325~MeV to 70~$\rm \mu$b at 400~MeV.

     Data are acquired in all eight possible combinations of beam (up,
down) and target polarization (up, down, left, right). As is clearly visible
in Fig.~2, the total cross section depends on the polarization of the colliding
protons, and the spin-dependent cross section,  $\rm \Delta\sigma_T$ (defined as the cross
section with the spins of the colliding protons vertical and opposite, minus
the cross section with spins parallel) can be calculated in a straight-forward
way from the corresponding four yields. The resulting values for  $\rm
\Delta\sigma_T/\sigma_{tot}$
are listed in Table~1, and shown in Fig.~3.

     In this experiment, complete kinematic information is obtained for
each event. Thus, spin-dependent observables other than $\rm \Delta\sigma_T$ can also be
investigated. To demonstrate this, we sort the yields as a function of
$\rm \varphi_{\pi}$,
the azimuthal angle of the $\rm \pi^{\circ}$. The columns in Fig.~4 correspond to the four
bombarding energies, and the rows to different combinations $\rm R_i$ of the
spin-dependent yields (for instance, $\rm R_1$=(sum of yields with beam spin
up)/(sum of  yields with beam spin down)). The yields for different spin
states are normalized to the same integrated luminosity. To obtain the
plotted quantity, we calculate $\rm W_i = (R_i-1)/(R_i+1)$, and divide
by $\rm (P_{BT})^{1/2}$
for the first three rows, or $\rm P_{BT}$ for the last two rows. The 
last two rows reflect the spin correlation coefficient combinations 
$\rm S\equiv (A_{xx}+A_{yy})$, and $\rm D\equiv (A_{xx}-A_{yy})$,
while the first three rows are related to the beam and target
analyzing powers, $\rm A_y^B$ and
$\rm A_y^T$, namely $\rm a_B=A_y^BP_B/(P_{BT})^{1/2}$, and 
$\rm a_T=A_y^TP_T/(P_{BT})^{1/2}$. The latter are defined such that
the product $\rm a_B\cdot a_T$ no longer depends on beam and target 
polarization . The
solid lines in Fig.~4 represent a least square fit using the known 
$\rm \varphi_{\pi}$
dependence (listed to the right of Fig.~4), varying the four
parameters 
$\rm a_B$, $\rm a_T$, S, and D. The fact that we ignore all kinematic
variables, except  $\rm \varphi_{\pi}$ means
that we implicitly integrate over the variables of the two protons
{\it and the
pion polar angle}. The beam and target analyzing powers, integrated
over
  $\rm \theta_{\pi}$, are related by $\rm A_y^B=-A_y^T\equiv A_y$. 
The integrated analyzing power is then
obtained by calculating $\rm A_y=(-a_B\cdot a_T)^{1/2}$. We note that the spin-dependent
total cross section is related to the integrated spin correlation coefficients
by  $\rm \Delta\sigma_T/\sigma_{tot} = -S$. This quantity can be deduced more directly as described
earlier, and the two analysis methods agree. The resulting polarization
observables are listed in Table~1, and plotted in Fig.~3.

     Our detector does not cover completely all of the phase space of the
exit channel because of the central hole that accommodates the circulating
beam, excluding particles with  $\rm \theta_{lab}\leq 5^{\circ}$, and causing a sizeable loss of
events (about 30\% at the lowest energy). This deficiency in acceptance was
studied by artificially enlarging the hole, accepting only events with both
protons outside a cone of opening angle $\rm \Theta_{hole}$. 
The resulting dependence of
 $\rm \Delta\sigma_T/\sigma_{tot}$ on $\rm \Theta_{hole}$ is weak, and consistent with an estimate based on the
difference in the geometrical acceptance for different exit channels. This
difference is caused by the final state interaction in the Ss channel (see
Ref.\cite{R01}) which enhances the part of phase space with protons of low
relative kinetic energy, and which is absent in the Ps and Pp channels. This
study yields a correction to  $\rm \Delta\sigma_T/\sigma_{tot}$ of the size of about one standard
deviation. Other possible systematic effects were found to be negligible.
They include a difference between the up and the down polarization of the
beam, a dependence of the acquisition dead time on the spin state, and
reasonable variations of the selection criteria for valid events. The validity
of the background subtraction is tested by varying the range of accepted
masses $\rm m_x$. A change of  the subtracted background by $\pm$25\% is used to
estimate an error which is added in quadrature to the statistical
errors, resulting in the errors listed in Table~1.

     Polarization observables permit a detailed study of individual partial
waves. In a first attempt to demonstrate this, we express the
observables near threshold in terms of ``partial wave'' contributions,
\begin{eqnarray}
 \Delta \sigma_T /\sigma_{tot} &=& -2 + (U/\sigma_{tot}) \eta^{\alpha} +
 (V_1/\sigma_{tot}) \eta^{\beta}\\ 
A_{xx} -A_{yy} &=& (V_2/\sigma_{tot}) \eta^{\beta}\\
A_y &=& (W/\sigma_{tot}) \eta^{(\alpha +\beta)/2}
\end{eqnarray}
where U and $\rm V_{1,2}$ correspond to transitions to a Ps or a Pp final state,
respectively, and W is due to interference between Ps and Pp. The
unpolarized total cross section is from the literature. The energy
dependence is described by a power law in 
$\rm \eta =p_{\pi,max}/m_{\pi}$. 
The dotted lines
in Fig.~3 are obtained with $\rm U=98~\mu$b, $\rm V_1=105~\mu$b, 
$\rm V_2=32~\mu$b, 
$\rm W=-22~\mu$b,
$\rm \alpha=4.8$, and  $\rm \beta=8$. The energy dependence of 
the Pp transitions is well represented by $\rm \beta=8$ (expected 
from phase space) but the Ps transitions show an energy
dependence ($\rm \alpha$) that is weaker than the expected power of 6. Note, that at
threshold ($\rm \eta=0$),  $\rm \Delta\sigma_T/\sigma_{tot}$ attains
the theoretical limit of $-$2.

     Most of the theoretical work on $\rm pp\rightarrow
     pp\pi^{\circ}$ deals with the lowest partial
wave (Ss). In fact, we are aware of only one calculation for
$\rm \Delta\sigma_T$  that
includes higher partial waves\cite{R12}, the result of which is shown
as a dashed
line in Fig.~3. This calculation includes the exchange of heavy mesons, and
provides a good fit to the total cross section below  $\rm \eta=0.5$
where the Ss final state dominates.

     Here, we have reported first results from a study of spin dependence
in $\rm \vec{p}\vec{p}\rightarrow pp\pi^{\circ}$. More information will be obtained from an analysis in terms
of different kinematic variables of the three-body final state, including
angular distributions. Moreover, a measurement that 
involves {\it longitudinal}
beam and target polarization is currently in progress. If history repeats
itself this new and detailed information on one of the basic processes in
nuclear physics will prove as stimulating to the theoretical community as
the earlier total cross section measurement\cite{R01}.

\begin{acknowledgements}
We would like to thank the accelerator operations group for their efforts.
We are also grateful to Ch. Hanhart for making the numerical values for the
calculation shown in Fig.~3 available. This work has been supported by the
US National Science Foundation under Grants PHY96-02872,  PHY95-
14566, PHY97-22556, and by the US Department of Energy under Grant
DOE-FG02-88ER40438.
\end{acknowledgements}

\begin{table}
\caption{Product of beam and target polarization, integrated luminosity, and the
three measured polarization observables as a function of bombarding
energy T, or $\rm \eta$ (maximum pion center-of-mass momentum divided by the
pion mass). The observables in the last two columns are for a detected
$\rm \pi^\circ$,
i.e.,  integrated over the variables of the two protons and the pion polar
angle.}
\label{table1}
\begin{tabular}{cccccrc}
T &   $\rm \eta$  &  $\rm P_B\cdot P_T$  &  $\rm \int$Ldt &
$\rm \Delta\sigma_T/\sigma_{tot}$, or  &
$\rm A_{xx}-A_{yy}$\enspace   &  $\rm A_y$  \\
 (MeV) & & & ($\rm \mu b)^{-1}$ & $\rm -(A_{xx}+A_{yy})$ & & \\ \tableline

325.6&0.560&0.456$\pm$0.003&2163&$-$1.038$\pm$0.075&$-$0.010$\pm$0.069&$-$0.066$\pm$0.018\\
350.5&0.707&0.342$\pm$0.004& 901&$-$0.450$\pm$0.086&   0.154$\pm$0.070&$-$0.160$\pm$0.023\\
375.0&0.832&0.514$\pm$0.004&3024&$-$0.285$\pm$0.021&   0.230$\pm$0.023&$-$0.169$\pm$0.006\\
400.0&0.948&0.526$\pm$0.006& 831&$-$0.070$\pm$0.046&   0.245$\pm$0.030&$-$0.224$\pm$0.008
\end{tabular}
\end{table}

\pagebreak

\input{epsf}
\epsfverbosetrue
\tolerance=1000
\baselineskip=14pt plus 1pt minus 1pt

\begin{figure}
\epsfxsize=8.5cm
\centerline{
\epsfbox{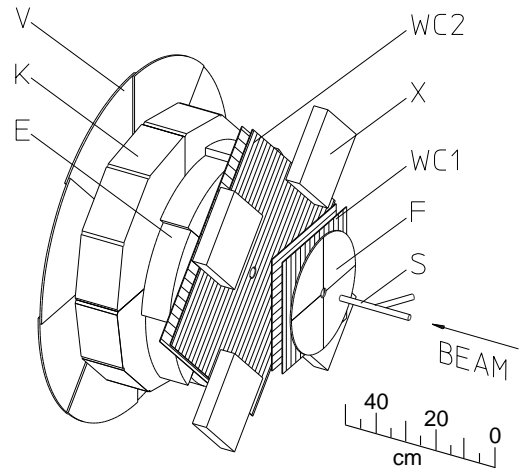}
}
\vspace{-0.5cm}
\caption{Detector setup and storage cell target. The components are explained
          in the text.}
\end{figure}

\pagebreak

\begin{figure}
\epsfxsize=8.5cm
\centerline{
\epsfbox{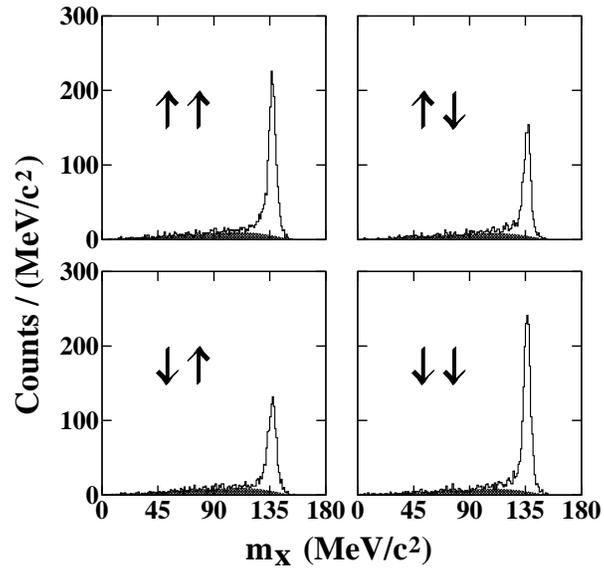}
}
\vspace{0.5cm}
\caption{Distributions of the calculated mass $\rm m_x$ of the undetected
          particle at 325~MeV bombarding energy, shown for the four
          combinations of vertical beam and target
          polarization. The peak of about  $\rm 7~MeV/c^2$ FWHM occurs at the
          $\pi^{\circ}$ rest mass. The shaded region indicates the
          assumed background distribution. }
\end{figure}

\begin{figure}
\epsfxsize=8.5cm
\centerline{
\epsfbox{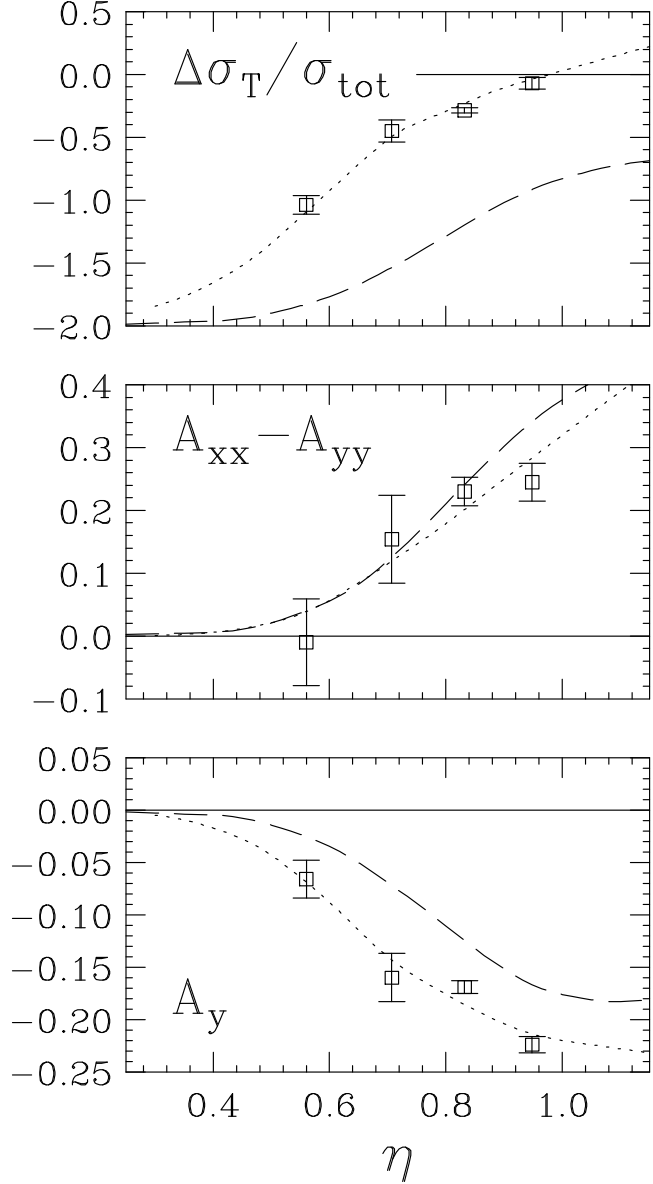}
}
\vspace{0.5cm}
\caption{The $\rm \vec{p}\vec{p}\rightarrow pp\pi^{\circ}$
  polarization observables  
$\rm \Delta\sigma_T/\sigma_{tot} = -(A_{xx}+A_{yy}$),
          ($\rm A_{xx}-A_{yy}$) and $\rm A_y$ as listed in Table~1 versus
$\rm \eta$, the maximum
          pion center-of-mass momentum in units of the pion mass. The 
          observables $\rm A_{xx}$, $\rm A_{yy}$ and $\rm A_{y}$ are
          integrated over the variables of the two protons and the
          pion polar angle. The
          dashed lines represent a meson-exchange calculation [12], and
          the dotted line is a partial-wave fit (explained in the text).}
\end{figure}

\pagebreak

\begin{figure}
\epsfysize=22.0cm
\centerline{
\epsfbox{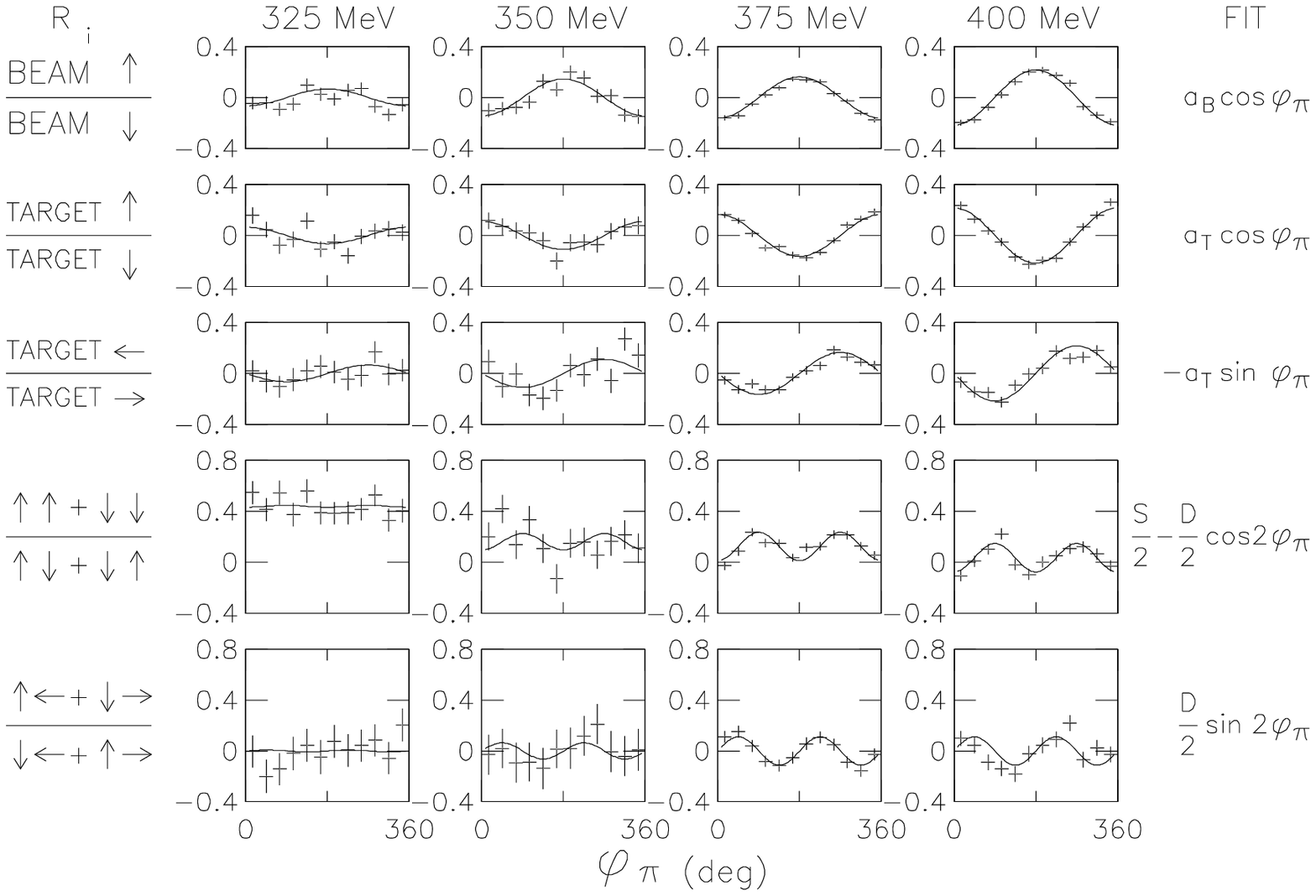}
}
\vspace{-4.0cm}
\caption{Asymmetries for different spin combinations $\rm R_i$ (listed on the
          left) as a function of the azimuthal angle of the
$\rm \pi^{\circ}$. The solid
          curves are obtained from a least-square fit using the theoretical
       $\rm \varphi_{\pi}$ dependence (listed on the right), varying $\rm
       a_B$,
$\rm a_T$, $\rm S\equiv (A_{xx}+A_{yy})$, and $\rm D\equiv (A_{xx}-A_{yy})$.}
\end{figure}


\begin{thebibliography}{12}
\bibitem{R01}  H.O. Meyer et al., Nucl.\ Phys.\ A{\bf 539}, 633 (1992)
  \bibitem{R02}A. Bondar et al., Phys.\ Lett.\ B{\bf 356}, 8 (1995)
\bibitem{R03}  H.O. Meyer, Annu.\ Rev.\ Nucl.\ Part.\ Sci. {\bf 47}, 235 (1997)
\bibitem{R04}  D.S. Koltun and A. Reitan, Phys.\ Rev.\ {\bf 141} (1966) 1413
\bibitem{R05}  E. Hernandez and E. Oset, Phys.\ Lett.\ B{\bf 350}, 158 (1995)
\bibitem{R06}  T.S.H. Lee and D.O. Riska, Phys.\ Rev.\ Lett. {\bf 70}, 2237 (1993)
\bibitem{R07}T. Sato, T.S.H. Lee, F. Myhrer and K. Kubodera, Phys.\ Rev.\ C{\bf 56}, 1246 (1997)
\bibitem{R08}  S. Stanislaus et al., Phys.\ Rev.\ C{\bf 44}, 2287 (1991)
\bibitem{R09}  G. Rappenecker et al., Nucl.\ Phys.\ A{\bf 590}, 763 (1995)
\bibitem{R10} W. Haeberli et al., Phys.\ Rev.\ C{\bf 55}, 597 (1997)
\bibitem{R11} B. von Przewoski et al., Phys.\ Rev.\ C, in print
\bibitem{R12} J. Haidenbauer, Ch. Hanhart and J. Speth, Acta\ Phys.\
Polonica\  B{\bf 27}, 2893 (1996)
\end{thebibliography}
\end{document}